\documentclass[fleqn,usenatbib]{mnras}
\usepackage{newtxtext,newtxmath}
\usepackage[normalem]{ulem}
\usepackage[T1]{fontenc}
\DeclareRobustCommand{\VAN}[3]{#2}
\let\VANthebibliography\thebibliography
\def\thebibliography{\DeclareRobustCommand{\VAN}[3]{##3}\VANthebibliography}
\usepackage{graphicx}	
\usepackage{amsmath}	
\usepackage{mathastext}
\usepackage{xcolor}

\title[A Theoretical Study of the Structure \& Elemental Abundances of HD~20794]{A Theoretical Study of the Structure and Elemental Abundances of HD~20794}

\author[M. Medhi et al. 2025]{
Mrinmay Medhi,$^{1}$\thanks{E-mail: mrinmaymedhi@gmail.com}
Mami Deka$,^{2}$
Krishna Saha,$^{1}$
Vivek Baruah Thapa,$^{3}$
and Upakul Mahanta$^{1}$
\\
$^{1}$Department of Physics, Bhattadev University, Pathsala, 781325, India\\
$^{2}$INAF -- Osservatorio Astronomico di Capodimonte,
Salita Moiariello, 16, 80131 Napoli NA, Italy\\
$^{3}$Department of Physics, Birangana Sati Sadhani Rajyik Vishwavidyalaya, Golaghat, 785621, India
}

\date{Accepted XXX. Received YYY; in original form ZZZ}

\pubyear{\the\year{}}

\begin{document}
\label{firstpage}
\pagerange{\pageref{firstpage}--\pageref{lastpage}}
\maketitle

\begin{abstract}
HD~20794 is a nearby, bright, and well-studied metal-poor G-type dwarf hosting a compact multi-planet system, including a super-Earth located near the habitable zone. Its low stellar activity and the availability of high-precision radial-velocity and photometric data make it a valuable benchmark for testing stellar structure models and chemical abundance interpretations in low-metallicity planet-hosting stars. To our best knowledge, we present the first grid-based stellar modelling analysis of HD 20794 using the Modules for Experiments in Stellar Astrophysics (\texttt{MESA}), focusing on its main-sequence and late main-sequence evolution. We computed 252 stellar evolutionary models with initial masses between 0.78 and 0.80~$M_\odot$, systematically varying the convection efficiency, numerical resolution, and atmospheric boundary conditions. We selected the models by performing a $\chi^2$-based comparison with the observed effective temperature, surface gravity, luminosity, radius, and age. The best-fitting models adopt a stellar mass of $0.80~M_\odot$ and an age of approximately 9~Gyr, and reproduce all observed stellar parameters within their uncertainties. In particular, the models successfully reproduce the observed surface abundance pattern of HD~20794 for a wide range of elements, including light elements, $\alpha$-elements, and the odd-$Z$ species phosphorus and chlorine. Comparisons with published massive-star nucleosynthesis yields suggest that the observed phosphorus and chlorine abundances are consistent with enrichment from core-collapse supernovae and have been preserved throughout the star’s evolution. {Our results are consistent with standard stellar evolution theory, indicating that low-mass, metal-poor G-type dwarfs such as HD~20794 retain their natal chemical signatures over Gyr timescales, providing valuable constraints on stellar evolution, Galactic chemical enrichment, and the chemical environments of long-lived planetary systems.}

\end{abstract}


\section{Introduction}
\label{sec:intro}
HD~20794 (82~G.~Eridani) is a nearby, bright G6V dwarf star ($V \approx 4.34$, distance $\sim 6.04$~pc) that has emerged as a key object in both stellar and exoplanetary studies \citep{feng2017}. Its fundamental stellar parameters are well constrained, with an effective temperature $T_{\rm eff} \approx 5368$~K, surface gravity $\log g \approx 4.39$, and a clearly sub-solar metallicity [Fe/H] $\approx -0.42$ \citep{feng2017,nari2025}. Combined spectroscopic and astrometric analyses indicate a mass of $\sim 0.79\,M_\odot$ and a radius of $\sim 0.93\,R_\odot$, making HD~20794 slightly cooler and smaller than the Sun \citep{nari2025}. Its high space motion is characteristic of thick-disk or old Galactic populations \citep{holmberg2007}, {with age estimates ranging from $\sim 5.8$~Gyr based on chromospheric activity \citep{sousa2015} to $\gtrsim 8$--10~Gyr from stellar modelling \citep{bonfanti2015,casagrande2011,nari2025}. Asteroseismic constraints are unavailable owing to the star’s intrinsically low oscillation amplitudes. Together, these properties identify HD~20794 as an old, metal-poor solar-type star and a valuable benchmark for stellar evolution studies in low-metallicity environments.}

HD~20794 also hosts a well-characterised system of low-mass planets detected through long-term high-resolution spectroscopy from High Accuracy Radial velocity Planet Searcher (HARPS) \citep{pepe2011,feng2017,nari2025}. 
{Early HARPS analyses suggested multiple planetary candidates, though initial interpretations were complicated by stellar activity \citep{pepe2011}. Subsequent studies employing improved radial-velocity (RV) extraction techniques \citep{feng2017,cretignier2021} resolved these ambiguities, leading to robust planet detections. A comprehensive synthesis by \citet{nari2025} reports three confirmed planets with orbital periods of 18.3, 89.7, and 647.6~days and minimum masses of $2.15$, $2.98$, and $5.82\, M_\oplus$. The outermost planet, HD~20794~d, resides within the star's conservative habitable zone, with a semi-major axis of approximately $0.9-1.0$~au. Its minimum mass of $\sim 5.8\,M_\oplus$ is consistent with a rocky or super-Earth classification, making it a compelling target for future assessments of habitability and atmospheric characterisation \citep{nari2025}. Although photometric observations, including  Transiting Exoplanet Survey Satellite (TESS), have not revealed transits, the system's well-constrained orbital parameters provide essential input for modelling potential habitable environments.}

The low metallicity of HD~20794 places it among stars formed from gas enriched early in the history of the Galactic disk \citep{matteucci2011}. Stars of this population typically exhibit distinct kinematic and chemical properties and play a key role in tracing Galactic chemical evolution \citep{bensby2005}. Metallicity is known to influence both planet formation and planetary architectures, with metal-poor stars rarely hosting giant planets while remaining efficient at forming super-Earths and Neptune-mass planets \citep{fischer2005,johnson2010,mortier2013}. The architecture of the HD~20794 system, which lacks gas giants but hosts multiple low-mass planets, is consistent with these trends. Metallicity further affects stellar radiation environments and planetary surface conditions. Recent studies suggest that planets orbiting metal-poor stars may experience reduced biologically damaging ultraviolet radiation due to the combined effects of stellar spectra and atmospheric ozone formation \citep{kriskovics2023}. Consequently, HD~20794 provides a unique laboratory for examining the interplay between stellar evolution, chemical composition, and planetary habitability in metal-poor environments.

The Galactic origin of phosphorus and chlorine remains an open problem in chemical evolution studies. Phosphorus is an odd-$Z$ element whose production is sensitive to neutron excess and explosion physics, and current Galactic chemical evolution models systematically underpredict the observed [P/Fe] trend in metal-poor and solar-metallicity stars \citep{caffau2011,maas2017,maas2022}. Similarly, the nucleosynthetic origin of chlorine is uncertain, with contributions proposed from hydrostatic oxygen burning, explosive neon burning, and neutrino-induced reactions in core-collapse supernovae, yet model predictions exhibit large dispersion and strong progenitor-mass dependence \citep{kobayashi2020,ritter2018}. Observational constraints on both elements remain sparse due to the weakness and blending of their diagnostic spectral lines, leading to significant uncertainties in abundance measurements \citep{caffau2011,cretignier2023}. As a result, it is unclear whether the measured P and Cl abundances in long-lived, low-mass stars reflect systematic model deficiencies, stochastic enrichment events, or internal stellar evolution effects. Establishing whether such abundances are preserved during stellar evolution is therefore a necessary prerequisite for interpreting them within a Galactic chemical evolution framework.

In this context, a detailed stellar evolution analysis is essential to determine whether the observed surface abundances of HD~20794 primarily reflect its natal chemical composition—imprinted by early Galactic enrichment—or have been altered during subsequent stellar evolution. Stellar evolution theory predicts that low-mass, main-sequence stars experience only minimal surface abundance changes over Gyr timescales, apart from modest effects due to microscopic diffusion and shallow envelope mixing \citep{dotter2007}. 
Despite extensive spectroscopic and planetary characterisation \citep[e.g.][]{pepe2011,mortier2013,nari2025}, HD~20794 has not previously been subjected to a systematic, grid-based stellar evolution analysis using modern one-dimensional stellar evolution codes that explicitly track the evolution of surface abundances. In particular, existing studies have focused primarily on fundamental stellar parameters and planetary architecture, without assessing whether the observed elemental abundances can be self-consistently preserved within a calibrated stellar structure model. In this work, we employ \texttt{MESA} stellar evolution models to constrain the internal structure and evolutionary state of HD~20794, and to test the stability of its surface chemical composition over its main-sequence lifetime. Particular attention is given to the odd-$Z$ elements phosphorus and chlorine, whose Galactic origins remain uncertain and observationally challenging \citep{maas2017,maas2022,ritter2018}. By combining stellar evolution modelling with comparisons to massive-star nucleosynthesis yields, we assess whether the observed phosphorus and chlorine abundances of HD~20794 are consistent with inheritance from earlier generations of core-collapse supernovae.


This paper is organised as follows. Section~\ref{sec:metho} presents the stellar evolution models computed with \texttt{MESA}, including the model grid, numerical setup, and calibration procedure, and analyses the sensitivity of the stellar parameters to variations in mass, convection efficiency, and numerical resolution, and identifies the best-fitting models through a $\chi^2$-based selection. Section~\ref{sec:abund} discusses the resulting surface abundance pattern and its origin, including a comparison with massive-star nucleosynthesis yields. Finally, Section~\ref{sec:diss} summarises our main conclusions and their implications for stellar evolution and chemical enrichment in metal-poor planet-hosting stars.

\begin{table}
\centering
\caption{Observed properties of HD~20794 and its planetary system.}
\label{tab:obs_constraints}
\begin{tabular}{lcc}
\hline
Parameter & Value & Reference \\
\hline
\multicolumn{3}{c}{\textbf{Stellar Parameters}} \\
Spectral type & G6V & \cite{nari2025} \\
$V$ (mag) & 4.34 & \cite{gaiadr3} \\
Distance (pc) & $6.04 \pm 0.01$ & \cite{gaiadr3} \\
$T_{\mathrm{eff}}$ (K) & $5368 \pm 64$ & \cite{nari2025} \\
$\log g$ (cgs) & $4.39 \pm 0.03$ & \cite{nari2025} \\
$L/L_\odot$ & $0.6869 \pm 0.0026$ & \cite{nari2025} \\
$R/R_\odot$ & $0.93 \pm 0.03$ & \cite{nari2025} \\
$M/M_\odot$ & $\sim0.79$ & \cite{nari2025} \\
$[\mathrm{Fe/H}]$ & $-0.42 \pm 0.02$ & \cite{nari2025} \\
Rotation period (d) & $\sim39$ & \cite{cretignier2023} \\
Age (Gyr) & $8$ -- $10$ & \cite{casagrande2011} \\
$\log R'_{\mathrm{HK}}$ & $-4.98$ & \cite{sousa2015} \\
\hline
\multicolumn{3}{c}{\textbf{Surface Abundances}} \\
$\epsilon(^{31}\mathrm{P})$ & $\sim5.37$ & \cite{maas2017} \\
$\epsilon(^{32}\mathrm{S})$ & $\sim6.65$ & \cite{maas2017} \\
$\epsilon(^{35}\mathrm{Cl})$ & $\sim4.73$ & \cite{nari2025} \\
\hline
\end{tabular}
\end{table}

\begin{figure*}
\centering
\includegraphics[width=0.8\textwidth]{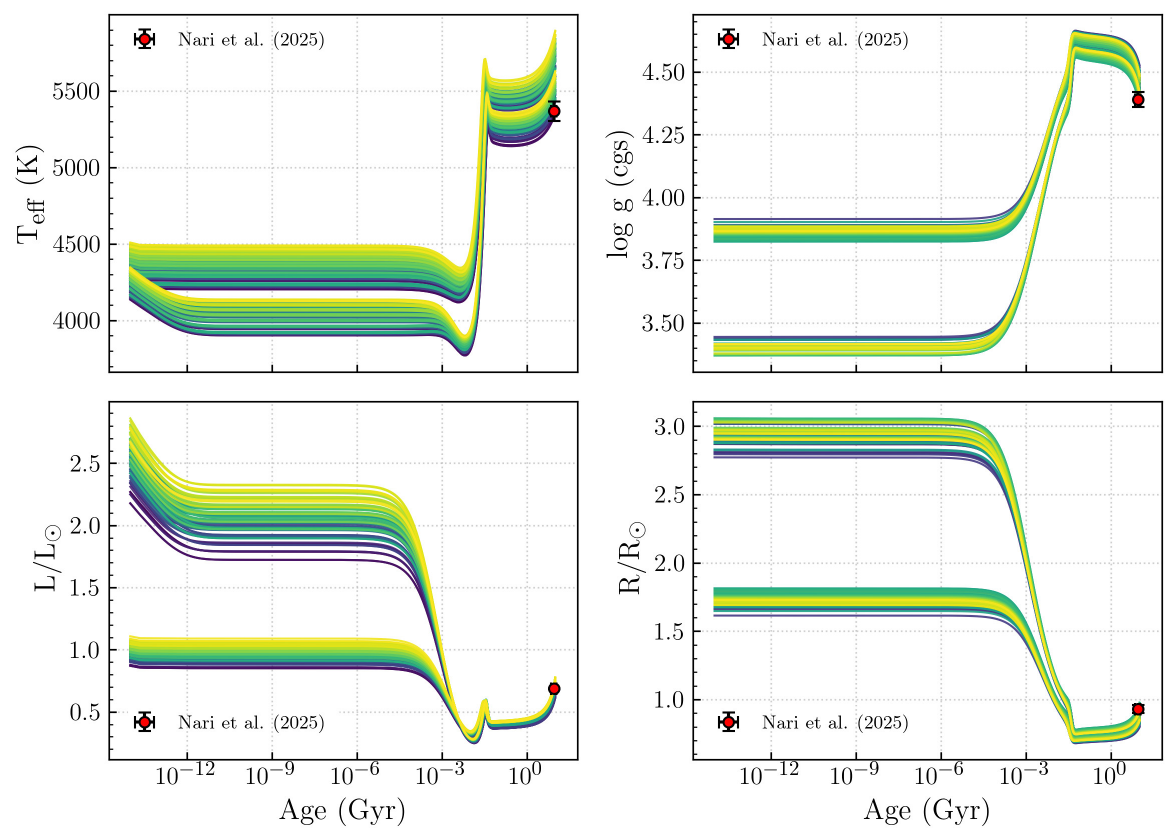}
\caption{Evolutionary tracks of HD~20794 computed with \texttt{MESA}, showing the time evolution of  effective temperature ($T_{\rm eff}$), surface gravity ($\log g$), luminosity ($L/L_\odot$), and radius ($R/R_\odot$) as a function of stellar age. {All models were computed assuming an initial metallicity $Z \simeq 0.005$ and helium abundances in the range $Y = 0.2525$--$0.2575$, corresponding to $\Delta Y/\Delta Z = 1.5$--$2.5$ at fixed metallicity.} Coloured curves represent models spanning the explored parameter grid in mass, mixing-length parameter, numerical resolution, and atmospheric boundary conditions. The red dot denotes the observed values from \citep{nari2025}. Error bars indicate the quoted observational uncertainties, which are smaller than the plotting symbols in some panels. Only a narrow subset of models simultaneously reproduces all four observables at a common age of $\sim$8--10~Gyr, supporting the adopted stellar parameters and evolutionary state of HD~20794. The apparent termination of the evolutionary tracks near the observed location reflects the fact that all models were evolved up to an age of 10~Gyr. The convergence of the tracks, therefore, indicates successful calibration rather than a physical endpoint of stellar evolution.}
\label{fig:evot}
\end{figure*}

\section{Methodology \& Analysis}
\label{sec:metho}
\subsection{Stellar Evolution Modelling with \texttt{MESA}}
We performed the stellar evolution calculations using the Modules for Experiments in Stellar Astrophysics (\texttt{MESA}) code \citep{paxton2010,paxton2013,paxton2015,paxton2018,paxton2019}, which solves the one-dimensional, time-dependent equations of stellar structure and composition under the assumptions of hydrostatic equilibrium and spherical symmetry. The models include standard microphysics as implemented in \texttt{MESA}, including OPAL Type II opacities, the standard MESA equation of state \citep{paxton2010,paxton2013,paxton2019}, which is a blend of the OPAL \citep{Rogers2002}, SCVH \citep{Saumon1995}, FreeEOS \citep{irwin2012}, HELM \citep{Timmes2000}, PC \citep{Potekhin2010}, and Skye \citep{Jermyn2021} EOSes and nuclear energy generation computed using an extended reaction network comprising the pp-chain and CNO cycles, with additional isotopes up to the $P-Cl$ region to enable tracking of odd-$Z$ elements. For a $0.8\,M_\odot$ star, the energy production is dominated by the pp-chain throughout the main-sequence evolution.

Given the sub-solar metallicity and low mass of HD~20794 ($M \simeq 0.79\,M_\odot$), energy production is almost entirely driven by the pp-chain, with only a negligible contribution from the CNO cycle. Reduced metallicity lowers the Rosseland mean opacity, leading to a more compact stellar structure relative to solar-metallicity stars. These effects are treated self-consistently within the \texttt{MESA} framework.

{Convective energy transport is treated using standard mixing-length theory \citep{bohmvitense1989}. Motivated by three-dimensional convection simulations that predict a metallicity dependence of the mixing-length parameter \citep{trampedach2013,magic2014}, we explored a broad range $\alpha_{\rm MLT}=1.2-1.8$, encompassing values both below and above the solar-calibrated value. Within this interval, variations in $\alpha_{\rm MLT}$ produce systematic shifts in $T_{\rm eff}$ and $R$. The minimum $\chi^2$ solutions lie in the range $\alpha_{\rm MLT}=1.4-1.6$, well within the explored parameter space and not at a grid boundary. We therefore restrict the final model grid to this narrower interval, which is consistent with both the observational constraints and theoretical expectations.}


We controlled numerical resolution using \texttt{MESA}'s adaptive mesh refinement by varying the mesh-delta coefficient over the range $0.1-2.0$. Unlike the mixing-length parameter, the mesh-delta coefficient has no physical meaning and primarily affects the numerical resolution of spatial gradients. For HD~20794, which remains on the main sequence and late main sequence throughout the evolution considered here, the stellar interior does not develop sharp composition or temperature gradients. As a result, the global stellar structure and surface observables are largely insensitive to the exact choice of mesh resolution. We verified numerical convergence by confirming that variations in effective temperature, luminosity, radius, and surface gravity across this range remain well below the observational uncertainties. The wide interval $0.1-2.0$ therefore yields effectively identical physical results, while allowing a balance between computational efficiency and numerical stability. Accordingly, we adopt this range as a conservative and converged choice for the model grid.

{Convective core overshooting was not included in the baseline calculations. For a star of mass $M \simeq 0.8\,M_\odot$, the core remains radiative throughout the main-sequence phase, and no convective core develops \citep{Kippenhahn2012,Salaris2005}. Core overshooting is therefore not physically relevant in this mass regime. We performed additional tests, including mild envelope overshooting using an exponential prescription ($f_{ov} = 0.01-0.02$). At 9 Gyr, the resulting variations in $T_{\rm eff}$, $L$, $R$, and $\log g$ remain significantly smaller than the observational uncertainties (Appendix~\ref{sec:overshoot}), confirming that overshooting does not significantly affect the inferred evolutionary state of HD~20794.}

The connection between the stellar interior and observable surface properties was established through atmospheric boundary conditions. We tested both a photospheric boundary at $\tau \approx 2/3$ and a deeper boundary at $\tau = 100$, where the diffusion approximation is valid. For HD~20794, models employing the $\tau_{100}$ boundary condition yield radii and effective temperatures in better agreement with observations, and this prescription is adopted for the preferred models.

{The importance of predictive stellar modelling and sensitivity to mixing prescriptions has been emphasised by \citet{young2001,young2005}, who analysed eclipsing binaries spanning $\sim 1.1-23\,M_\odot$ and demonstrated that incomplete treatment of mixing can lead to systematic discrepancies, particularly in the $1.7-2.6\,M_\odot$ regime where convective cores are present. Their quantitative constraints on mixing and overshooting are therefore calibrated for stars with $M \gtrsim 1.1\,M_\odot$. In contrast, HD~20794 ($M \simeq 0.79\,M_\odot$) lies below this mass range and retains a radiative core throughout the main-sequence phase. Accordingly, we restrict our models to a physically motivated approach for this mass regime and assess robustness through explicit parameter-sensitivity and convergence tests within the \texttt{MESA} framework.}


\subsection{Phosphorus and Chlorine as Passive Tracers}
{In addition to the standard pp-chain and CNO reactions, we employed a custom PCl nuclear network implemented within \texttt{MESA} and described in \citet{medhi2023}, which extends the hydrogen-burning network to include isotopes up to phosphorus and chlorine. For a $0.79\,M_\odot$ main-sequence star such as HD~20794, the central temperatures remain well below those required to activate proton-capture cycles involving these species. Consequently, phosphorus and chlorine are not modified by in-situ nucleosynthesis during the star’s evolution. In this mass regime, P and Cl therefore behave as passive tracers of the natal chemical composition. Their inclusion allows us to verify abundance preservation self-consistently and to compare the resulting surface abundances directly with spectroscopic measurements and theoretical nucleosynthesis yields.}

\subsection{Model Grid Consideration}
A grid of 252 stellar evolution models was computed to sample the parameter space relevant to HD~20794. The grid was constructed by varying the following input quantities:

\begin{enumerate}
    \item {Stellar mass:} $M = 0.78$--$0.80\,M_\odot$ in steps of $0.01\,M_\odot$
    \item {Mixing-length parameter:} $\alpha_{\mathrm{MLT}} = 1.2-1.8$ in steps of $0.1$
    \item {Mesh refinement coefficient:} $\Delta_{\mathrm{mesh}} = 0.1-2.0$ in steps of $1.0$
    \item {Initial metallicity:} $Z = 0.0045-0.0050$ in steps of $5\times10^{-4}$
    \item {Atmospheric boundary conditions:} photospheric and $\tau_{100}$
\end{enumerate}

Each model was initialised from a fully relaxed zero-age main-sequence (ZAMS) configuration and evolved up to 10 Gyr. Figure~\ref{fig:evot} presents the resulting evolutionary tracks of HD~20794 computed with \texttt{MESA}, showing the time evolution of effective temperature, surface gravity, luminosity, and radius. The coloured curves correspond to models spanning the adopted grid in stellar mass, mixing-length parameter, and atmospheric boundary conditions, while the red dot indicates the observed stellar parameters reported in the literature. These tracks define the region of parameter space explored in this study and provide the basis for the quantitative comparison discussed in Section~\ref{sec:diss}.

\subsection{Numerical Controls and Convergence}
To ensure numerical stability and robust convergence, the stellar evolution equations were solved using \texttt{MESA}’s differential energy formulation together with the recommended ``gold tolerances,'' which enforce stricter convergence criteria in the Newton iteration scheme \citep{paxton2010,paxton2013}. Adaptive time-stepping and adaptive mesh refinement were employed throughout the model grid.{ We explicitly verified that during the late main-sequence phase (8$-$10 Gyr), the timestep remains below $\sim 8$~yr, with a mean value of $\sim 6$~yr. This corresponds to more than $10^8$ temporal resolution elements over the stellar lifetime and ensures that the inferred stellar parameters are insensitive to timestep size.}

{To ensure numerical robustness, we performed explicit convergence tests prior to adopting the final model grid. The spatial resolution was varied by changing the mesh refinement parameter between 1.0 and 2.0 while keeping all physical parameters fixed. Independent MCMC analyses were conducted for each resolution level. The resulting posterior medians differ by 
$\Delta T_{\rm eff} \approx 0.3$~K,
$\Delta L/L_\odot \approx 1.5\times10^{-5}$~dex,
$\Delta \log g \approx 5\times10^{-4}$~dex,
$\Delta R \approx 5\times10^{-4}\,R_\odot$, and
$\Delta {\rm Age} \approx 0.03$~Gyr. These differences are more than an order of magnitude smaller than the observational uncertainties. Therefore, the inferred stellar parameters are numerically converged and insensitive to the adopted mesh resolution within the explored range.}

\subsection{Parameter Sensitivity Analysis}
\label{sec:sensitivity}
{To quantify how stellar observables (age, $T_{\rm eff}$, $L$, $R$, and $\log g$) respond to changes in the input parameters (mass $M$, mixing-length parameter $\alpha_{\rm MLT}$, and mesh resolution), we analysed homology-based scaling relations together with regression trends extracted from our model grid. All derivatives quoted below refer to variations around the best-fitting region of parameter space, defined by $(M_0 = 0.80\,M_\odot,\ \alpha_{\rm MLT} = 1.4-1.6$, $Z = 0.0050)$.}

Within the explored parameter space, stellar mass provides the strongest control on luminosity and evolutionary timescale \citep{Kippenhahn2012}. A change of $\Delta M = 0.01\,M_\odot$ produces a luminosity shift of $\sim 0.02$ dex and a temperature variation of $\sim 40$ K. Variations in $\alpha_{\rm MLT}$ of $\Delta\alpha = 0.1$ produce comparable shifts in $T_{\rm eff}$ but negligible changes in luminosity. This near-degeneracy between mass and convection efficiency generates an elongated likelihood ridge in the $(M, \alpha_{\rm MLT})$ plane, while luminosity provides the strongest constraint for selecting the preferred models. A summary of the relative parameter sensitivities is provided in Table~\ref{tab:sensitivity}.


\begin{figure*}
\centering
\includegraphics[width=0.8\textwidth]{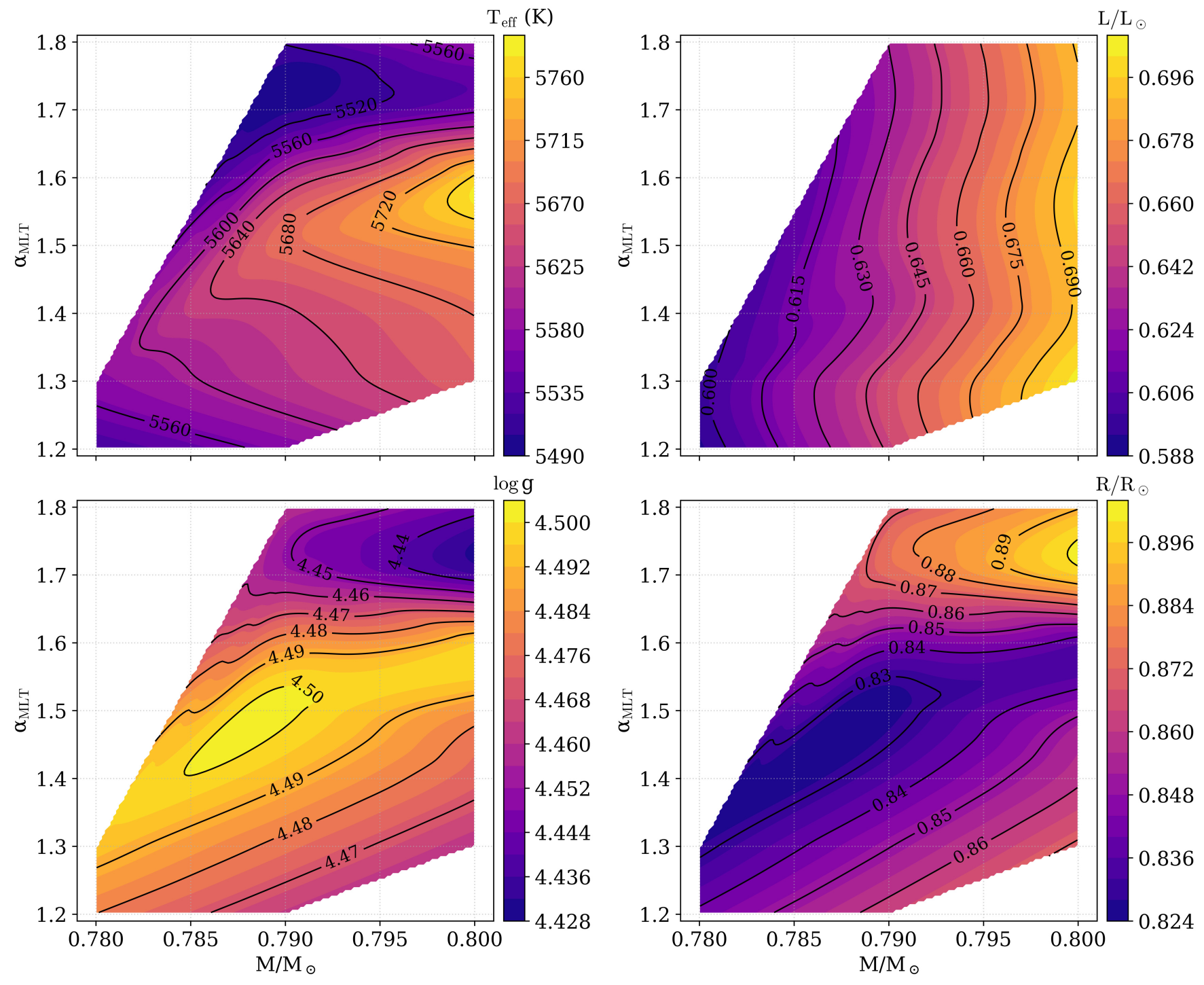}
\caption{Two-dimensional contour maps showing the dependence of
$\mathrm{T_{eff}}$, $\mathrm{L/L_\odot}$, $\log g$, and $\mathrm{R/R_\odot}$ on stellar mass and mixing-length parameter at an age of 9~Gyr for $Z=0.0050$.}
\label{fig:contourZ005}
\end{figure*}

\begin{table}
\centering
\caption{Sensitivity of stellar observables to parameter variations near the best-fit model.}
\label{tab:sensitivity}
\begin{tabular}{lccc}
\hline
Observable & $\Delta M = +0.01\,M_\odot$ & $\Delta \alpha = +0.1$ & Mesh variation \\
\hline
$T_{\rm eff}$ [K] & +42 & +40 & 10 -- 50 \\
$L/L_\odot$ [dex]    & +0.022 & 0.002 -- 0.01 & 0.005 -- 0.02 \\
$R/R_\odot$ [\%]  & +1.0\% & $-1.5$\% & 0.2 -- 2\% \\
$\log g$ [dex]    & $-0.0033$ & +0.03 & 0.01 -- 0.03 \\
\hline
\end{tabular}
\end{table}

Although the model grid spans a range of ages and metallicities, the structure of the mass -- $\alpha_{\rm MLT}$ degeneracy and the inferred best-fit parameters remain stable across the explored domain. Consequently, the results presented here are not sensitive to the specific choice of reference age or numerical resolution, but reflect robust physical trends inherent to the stellar models.

\begin{table*}
\centering
\caption{Best-fitting stellar models for HD~20794 selected from the
$\chi^2$ minimisation procedure. Only models with
$\chi^2_{\mathrm{red}} \leq 2$ are shown.}
\label{tab:bestfit}
\setlength{\tabcolsep}{6pt}
\begin{tabular}{cccccccccccc}
\hline
ID &
$M$ &
$\alpha_{\rm MLT}$ &
Mesh &
$\tau_{\rm atm}$ &
$Z$ &
$T_{\rm eff}$ &
$\log g$ &
$L/L_\odot$ &
$R/R_\odot$ &
Age &
$\chi^2_{\rm red}$ \\
 &
($M_\odot$) & & & & & (K) & (cgs) & & & (Gyr) & \\
\hline
1 & 0.80 & 1.4 & 2.0 & 100 & 0.005 &
5475 & 4.44 & 0.69 & 0.92 & 9.10 & 1.00 \\
2 & 0.80 & 1.4 & 1.0 & 100 & 0.005 &
5480 & 4.44 & 0.69 & 0.92 & 9.09 & 1.13 \\
3 & 0.80 & 1.6 & 2.0 & 100 & 0.005 &
5530 & 4.43 & 0.69 & 0.90 & 8.99 & 1.79 \\
\hline
\end{tabular}
\end{table*}

\subsection{Influence of Convection and Mesh Resolution}
Beyond stellar mass, the mixing-length parameter $\alpha_{\mathrm{MLT}}$ exerts a strong influence on the surface properties of the models. Within the parameter space relevant to HD~20794, increasing $\alpha_{\mathrm{MLT}}$ leads to systematically higher effective temperatures and smaller radii. Quantitatively, we find that $\Delta\alpha_{\mathrm{MLT}} = 1$ corresponds to an increase in $T_{\mathrm{eff}}$ of approximately $300$~K and a reduction in radius of $\sim 0.1\,R_\odot$. This behaviour is a direct consequence of mixing-length theory, whereby more efficient convection reduces the superadiabatic gradient in the outer envelope, producing hotter and more compact stellar structures.

In contrast, variations in the mesh-refinement coefficient affect only the numerical representation of the stellar structure. Across the explored range of mesh resolutions, the resulting changes in $T_{\mathrm{eff}}$, $L$, $R$, and $\log g$ remain well below the observational uncertainties and do not alter the inferred best-fitting parameters. As expected from its purely computational role, mesh resolution therefore has no impact on the physical interpretation of the models \citep{paxton2013}.

\subsection{Calibration Procedure}
To identify stellar models consistent with the observational constraints, we performed a $\chi^2$ minimisation using the measured stellar parameters:
effective temperature, surface gravity, luminosity, radius, and age. For each model in the grid, the goodness of fit is quantified as:
\begin{equation}
\chi^2 =
\sum_i
\frac{\left(X_{i,\mathrm{model}} - X_{i,\mathrm{obs}}\right)^2}{\sigma_i^2},
\end{equation}
where $X_i$ denotes the set of observables ($T_{\mathrm{eff}}$, $\log g$, $L$, $R$, and age) and $\sigma_i$ their corresponding uncertainties. The reduced chi-square, $\chi^2_{\mathrm{red}} = \chi^2/\nu$, where the number of degrees of freedom
$\nu = N_{\mathrm{obs}} = 5$, is computed using the appropriate number of degrees of freedom $\nu$. Models with $1 \lesssim \chi^2_{\mathrm{red}} \lesssim 2$ are considered acceptable fits. The minimum $\chi^2_{\mathrm{red}}$ is obtained for models with stellar mass $M \simeq 0.80\,M_\odot$, mixing-length parameter $\alpha_{\mathrm{MLT}} = 1.4-1.6$, with the global minimum located at $\alpha_{\rm MLT} = 1.4$ and metallicity $Z \simeq 0.0050$, at an age of approximately $9$~Gyr. 

These models simultaneously reproduce all observed stellar parameters within their uncertainties and are consistent with independent age estimates for HD~20794. The distribution of acceptable solutions reflects the intrinsic degeneracy between stellar mass and mixing-length parameter discussed in Section~\ref{sec:sensitivity}, rather than numerical uncertainty. Taken together, the $\chi^2$ calibration confirms that HD~20794 is a low-mass, metal-poor solar-type star approaching the terminal-age main sequence.

\subsection{Model Validation}
\begin{figure}
\centering
\includegraphics[width=0.45\textwidth]{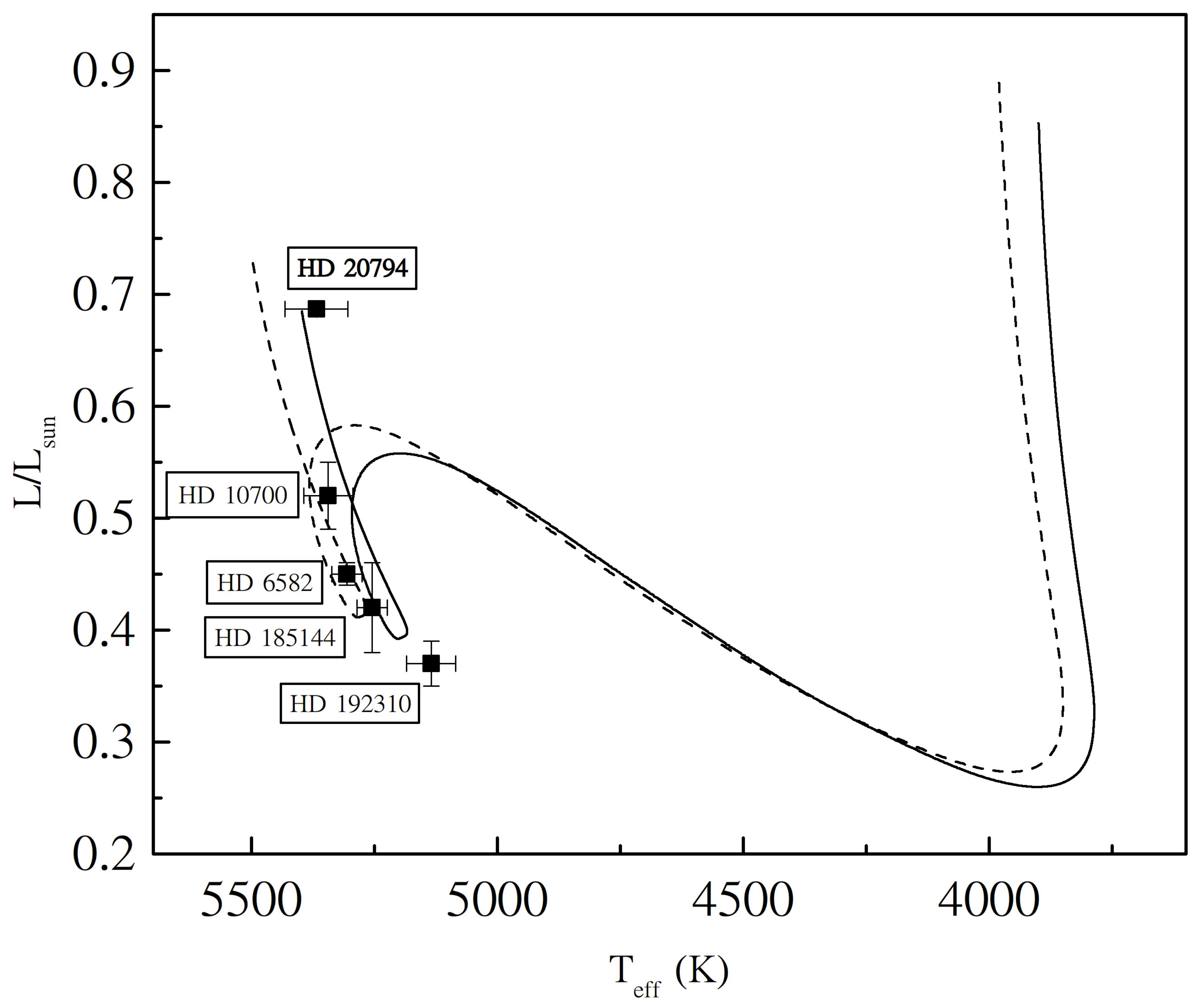}
\caption{HR diagram showing the location of HD~20794 (filled star) together with \texttt{MESA} evolutionary tracks computed for stellar masses and mixing-length parameters near the best-fit solution. Solid and dashed curves correspond to models spanning the preferred parameter range ($M \simeq 0.78$--$0.80\,M_\odot$, $\alpha_{\mathrm{MLT}} = 1.4$, $Z \simeq 0.0050$). For comparison, several nearby solar-type stars with well-determined parameters (HD~10700, HD~6582, HD~185144, and HD~192310) are also shown. The position of HD~20794 lies close to the terminal-age main sequence, consistent with an age of $\sim$9~Gyr inferred from the calibrated stellar models.}
\label{fig:ch6_HRDiagram}
\end{figure}

Figure~\ref{fig:ch6_HRDiagram} presents the location of HD~20794 in the Hertzsprung--Russell (HR) diagram together with the evolutionary tracks derived from our best-fit \texttt{MESA} models. The solid and dashed curves correspond to stellar masses and mixing-length parameters spanning the narrow region of parameter space favoured by the $\chi^2$ calibration. The observed position of HD~20794 lies well within the model uncertainties in effective temperature and luminosity, demonstrating that the adopted stellar parameters—most notably $M \simeq 0.80\,M_\odot$, $Z \simeq 0.0050$, and $\alpha_{\mathrm{MLT}} = 1.4$ successfully reproduce the star’s global properties.

For context, the figure also includes several nearby solar-type stars with well-determined parameters (HD~10700, HD~6582, HD~185144, and HD~192310). In comparison with these analogues, HD~20794 is slightly more evolved, occupying a position close to the terminal-age main sequence. This agreement between observations and theoretical tracks confirms the internal consistency of the stellar models and validates the physical assumptions adopted in the \texttt{MESA} simulations.

\section{Abundance Estimation}
\label{sec:abund}
To quantify the surface composition of HD~20794, we extracted the mass fractions and corresponding logarithmic abundances ($\epsilon_X$) from the best-fit stellar model at an age of approximately $9~\mathrm{Gyr}$. The abundances were computed using the Equation, $\epsilon_X = \log(N_X/N_H)+12$. The resulting mass fractions and $\epsilon$ values are listed in Table~\ref{tab:abundances_hd20794_main}. The literature comparison column ($\epsilon_{\mathrm{lit}}$) includes observed photospheric abundances for HD~20794 from spectroscopic analyses.{ The Solar abundances are adopted from \citet{asplund2021}. All logarithmic abundances and abundance ratios presented in this work are computed relative to this reference scale.}

\begin{table}
\centering
\caption{Surface mass fractions and logarithmic abundances of selected stable elements in the best-fit HD~20794 model (age $\simeq 9$~Gyr), compared with literature values. Only astrophysically relevant and observable species are listed. Complete surface abundance output from the extended nuclear network is presented in Table~\ref{tab:abundances_all_hd20794}.}
\label{tab:abundances_hd20794_main}
\begin{tabular}{clcccc}
\hline
No. & Element & Isotope & Mass Fraction & $\epsilon_{\rm model}$ & $\epsilon_{\rm lit}$ \\
\hline
1  & H  & $^{1}$H  & $7.42\times10^{-1}$ & 12.00 & 11.96 \\
2  & He & $^{4}$He & $2.55\times10^{-1}$ & 10.92 & 11.50 \\
3  & C  & $^{12}$C & $7.66\times10^{-4}$ & 7.93 & 8.43 \\
4  & N  & $^{14}$N & $2.26\times10^{-4}$ & 7.34 & 8.05 \\
5  & O  & $^{16}$O & $2.11\times10^{-3}$ & 8.25 & 8.77 \\
6  & Ne & $^{20}$Ne& $4.34\times10^{-4}$ & 7.46 & 8.05 \\
7  & Mg & $^{24}$Mg& $1.78\times10^{-4}$ & 7.00 & 7.54 \\
8  & P  & $^{31}$P & $2.19\times10^{-6}$ & 4.98 & 5.50 \\
9  & S  & $^{32}$S & $9.89\times10^{-5}$ & 6.62 & 7.10 \\
10 & Cl & $^{35}$Cl& $1.30\times10^{-6}$ & 4.70 & 5.38 \\
\hline
\end{tabular}
\end{table}

The derived surface abundance pattern is consistent with that of an old, mildly metal-poor solar-type star \citep{nari2025}.
Hydrogen and helium dominate the stellar envelope. To account for uncertainties in Galactic helium enrichment, we computed models spanning helium mass fractions $Y = 0.2525-0.2575$, corresponding to $\Delta Y/\Delta Z = 1.5-2.5$ at $Z \simeq 0.005$ (Appendix~\ref{sec:helium}). Across this range, the resulting variations in $T_{\rm eff}$, $L$, and $R$ at 9~Gyr remain within or comparable to the observational uncertainties and do not alter the inferred evolutionary state. Heavy elements contribute a total metallicity of $Z \simeq 0.0050$, consistent with the spectroscopically inferred [Fe/H] $\approx -0.4$ of HD~20794. Given the low mass of HD~20794, its main-sequence core temperatures remain insufficient to activate advanced $\alpha$-capture or heavy proton-capture reactions. The present-day surface abundances therefore largely reflect the star’s natal chemical composition, with only minor modifications arising from diffusion and envelope mixing.

The $\alpha$-elements (O, Mg, Si, and S) exhibit mildly enhanced ratios relative to iron, consistent with enrichment dominated by core-collapse supernovae in the early Galactic disc \citep{bensby2005}. Carbon, nitrogen, and oxygen follow the trends observed in metal-poor solar-type dwarfs, while the odd-$Z$ elements phosphorus and chlorine fall within the range reported for Galactic thin and thick-disk stars \citep{maas2022}. Although systematic offsets of a few tenths of a dex are present for individual elements, the relative abundance pattern is reproduced well by the calibrated stellar model.

Importantly, the agreement between model-predicted and observed abundances does not imply in-situ production within HD~20794. Instead, the inclusion of an extended nuclear network serves to track the evolution of these species as passive tracers of the initial composition. The consistency of the surface abundances therefore supports the interpretation that HD~20794 has preserved its natal chemical signature and is presently in a late main-sequence evolutionary state approaching the terminal-age main sequence \citep{nari2025}.

\subsection{Origin of the Surface Abundances}
\begin{figure*}
\centering
\includegraphics[width=0.8\textwidth]{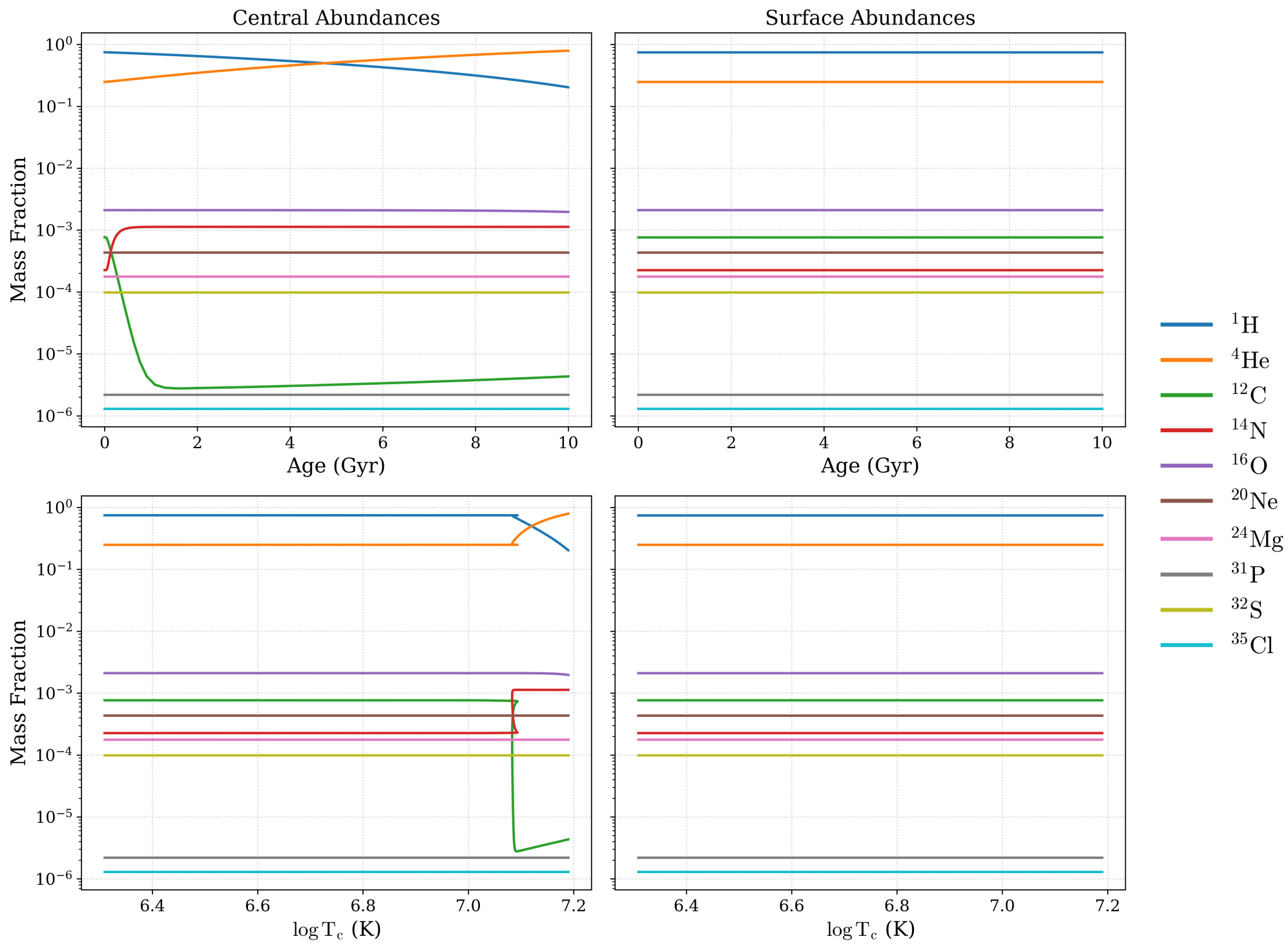}
\caption{Time evolution of central and surface mass fractions for selected elements in the best-fit \texttt{MESA} model of HD 20794.
\textit{Top panels:} Central (left) and surface (right) abundances as a function of stellar age. \textit{Bottom panels:} Central (left) and surface (right) abundances as a function of central temperature.}
\label{fig:abundace}
\end{figure*}

{For a low-mass main-sequence star such as HD~20794 ($M \simeq 0.8\,M_\odot$), stellar evolution theory predicts that surface abundances remain largely unchanged over gigayear timescales, apart from modest effects due to diffusion \citep{dotter2007}. Our \texttt{MESA} models are consistent with this expectation: the abundances of heavy elements, including phosphorus and chlorine, show no measurable variation throughout the $\sim 10$~Gyr evolution (Figure~\ref{fig:abundace}). 
These elements therefore, act as passive tracers of the initial composition.}

\subsubsection{Comparison with Massive-Star Nucleosynthesis Yields}
\label{sec:yields}
To investigate the origin of the birth abundances of HD~20794, we compared the observed elemental abundance ratios with published massive-star nucleosynthesis yields from the NuGrid collaboration \cite{ritter2018}. We adopted the NuGrid Set~1.1 models, corresponding to sub-solar metallicity ($Z \simeq 0.006$), which serve as an appropriate proxy for the metallicity of HD~20794 ($[\mathrm{Fe/H}] \approx -0.4$). Elemental yields of O, Si, P, and Cl were obtained by summing the relevant isotopic ejecta for each element and converting the resulting masses into spectroscopic abundance ratios relative to iron.

Table~\ref{tab:nugrid_xfe} lists the resulting NuGrid [$X$/Fe] ratios for progenitor masses between 12 and $25\,M_{\odot}$. The predicted abundance ratios exhibit a strong dependence on progenitor mass, with particularly large scatter for the odd-$Z$ elements phosphorus and chlorine. Models with progenitor masses of $15-20\,M_{\odot}$ yield [$X$/Fe] ratios that are close to solar or mildly enhanced, while the $25\,M_{\odot}$ model shows extreme enhancements driven by significantly reduced iron ejecta due to fallback. Such behaviour is a known feature of single-event core-collapse supernova yields and does not directly reflect the chemical composition of the well-mixed interstellar medium. Also, we restrict the analysis to massive-star models ($M \geq 12\,M_\odot$) and exclude lower-mass AGB models, which do not contribute significantly to iron production. To quantify the contribution of different progenitor masses, we computed initial mass function (IMF) weighted abundance ratios assuming a Salpeter IMF, $\phi(M)\propto M^{-2.35}$. The resulting IMF-averaged abundances for different mass intervals are summarized in Table~\ref{tab:nugrid_imf}. Including progenitors close to the minimum core-collapse mass (12~$M_\odot$) leads to sub-solar [$\alpha$/Fe] ratios, whereas restricting the analysis to progenitors in the mass range $15-20~M_\odot$ yields abundance ratios that are fully consistent with the observed chemical pattern of HD~20794.

Overall, the comparison with massive-star nucleosynthesis yields is intended to assess consistency rather than to identify a unique enrichment source. Individual core-collapse supernova models exhibit substantial scatter due to explosion energy and fallback, particularly for odd-$Z$ elements. When averaged over a Salpeter initial mass function and restricted to progenitor masses of $15-20$ $M_\odot$, the predicted abundance ratios of O, Si, P, and Cl closely match those observed in HD~20794. This supports a formation scenario in which the star inherited its chemical composition from gas predominantly enriched by mid-mass core-collapse supernovae, while our stellar evolution models demonstrate that these abundances are preserved throughout the subsequent main-sequence evolution.


\begin{table}
\centering
\caption{Individual abundance ratios [X/Fe] predicted by NuGrid massive-star models (Set~1.1, $Z=0.006$) \citep{ritter2018}.}
\label{tab:nugrid_xfe}
\begin{tabular}{ccccc}
\hline
$M$ ($M_\odot$) & [O/Fe] & [Si/Fe] & [P/Fe] & [Cl/Fe] \\
\hline
12 & $-0.63$ & $-0.03$ & $-0.10$ & $-0.19$ \\
15 & $+0.05$ & $+0.22$ & $+0.17$ & $-0.39$ \\
20 & $-0.00$ & $+0.35$ & $+0.32$ & $+0.09$ \\
25 & $+2.16$ & $+1.61$ & $+2.28$ & $+1.80$ \\
\hline
\end{tabular}
\end{table}

\begin{table}
\centering
\caption{IMF-weighted abundance ratios [X/Fe] computed from NuGrid yields assuming a Salpeter IMF.}
\label{tab:nugrid_imf}
\begin{tabular}{lcccc}
\hline
Mass range ($M_\odot$) & [O/Fe] & [Si/Fe] & [P/Fe] & [Cl/Fe] \\
\hline
12 -- 20 & $-0.20$ & $+0.13$ & $+0.08$ & $-0.18$ \\
\textbf{15 -- 20} & \textbf{+0.03} & \textbf{+0.27} & \textbf{+0.23} & \textbf{$-0.17$} \\
\hline
\end{tabular}
\end{table}

\section{Discussion and Conclusions}
\label{sec:diss}

{Accurate stellar modelling is essential for interpreting the observed properties of HD~20794 and for constraining its evolutionary state. In this work, we computed a grid of stellar evolutionary models using the \texttt{MESA} code, spanning masses $M = 0.78 - 0.80\,M_\odot$, mixing-length parameters $\alpha_{\mathrm{MLT}} = 1.4$, and metallicities $Z \simeq 0.0045-0.0050$, consistent with the observed [Fe/H] of HD~20794. By calibrating the models against spectroscopic and astrometric constraints ($T_{\rm eff}$, $\log g$, $L$, $R$), we identify a best-fit solution with a stellar mass of $M \simeq 0.80\,M_\odot$ and an age of $\sim 9$~Gyr, placing the star near the terminal-age main sequence. The fitted stellar parameters reproduce the observed location of HD~20794 in the HR diagram (Figure~\ref{fig:ch6_HRDiagram}), confirming the internal consistency of the adopted physics and numerical setup.}


Our sensitivity analysis shows that stellar mass provides the dominant contribution to variations in luminosity, radius, and effective temperature within the explored parameter range.
Variations in the mixing-length parameter predominantly affect surface properties, producing hotter and more compact models for larger $\alpha_{\mathrm{MLT}}$, while changes in mesh resolution have no measurable impact on global stellar observables. These results validate the robustness of the model grid and confirm that the inferred stellar parameters are not driven by numerical artefacts. In addition, we explored the impact of helium enrichment by varying the initial helium abundance in the range $Y = 0.2525$--$0.2575$, corresponding to $\Delta Y/\Delta Z = 1.5$--$2.5$ at fixed $Z \simeq 0.005$. At an age of 9~Gyr, the resulting shifts in $T_{\rm eff}$, $L$, and $R$ remain within or comparable to the observational uncertainties and do not alter the inferred evolutionary state. We further tested mild envelope overshooting using the exponential diffusive prescription with $f_{\rm ov}=0.01$ and $0.02$. The inclusion of overshooting produces changes in global parameters well below observational errors and does not modify our conclusions. 
These results indicate that the inferred stellar parameters are not significantly affected by numerical choices or specific assumptions about mixing.

The evolution of elemental abundances provides a consistency check for the stellar models.
The four-panel abundance evolution shown in Figure~\ref{fig:abundace} demonstrates that hydrogen is steadily depleted and helium correspondingly enhanced in the core via the proton -- proton chain, while the abundances of heavier elements remain effectively constant throughout the star’s $\sim 10$~Gyr evolution. The central temperature never exceeds $\log T_c \approx 7.2$, far below the threshold required to activate proton-capture cycles involving intermediate-mass nuclei. As a consequence, no PCl nucleosynthetic channels operate, and the surface abundances of C, O, Ne, Mg, P, S, and Cl preserve their natal values. The extended nuclear network thus acts as a diagnostic tracer rather than an active source of heavy-element production.

To place these results in a broader nucleosynthetic context, we compared the observed abundance ratios of HD~20794 with massive-star nucleosynthesis yields from the NuGrid collaboration at sub-solar metallicity. The observed [$X$/Fe] ratios for O, Si, P, and Cl are consistent with yields from core-collapse supernovae in the progenitor mass range $15-20\,M_\odot$, when averaged over a standard initial mass function. This agreement supports a formation scenario in which HD~20794 inherited its chemical composition from gas enriched primarily by earlier generations of massive stars, rather than through internal stellar nucleosynthesis.

In summary, HD~20794 is an old, metal-poor, solar-type star hosting a system of low-mass planets. By combining precise observational constraints with self-consistent, this study presents a grid-based stellar evolution analysis of HD~20794 using \texttt{MESA}, enabling a quantitative assessment of its evolutionary state and surface abundance preservation. The results show that the observed abundances of phosphorus and chlorine originate from Galactic chemical enrichment rather than in-situ stellar processes, while demonstrating that low-mass, metal-poor G dwarfs can serve as long-lived fossil records of early nucleosynthesis. 
This work illustrates how stellar evolution modelling combined with abundance tracking can aid the interpretation of chemically well-characterised exoplanet host stars.


\section*{Acknowledgements}
We thank J. L. Gutiérrez for the constructive and insightful comments, which have significantly improved the clarity and quality of this manuscript. The authors are also grateful to Kanak Saha for valuable discussions that contributed to the development of this work. MD acknowledges funding from the INAF 2023 Large Grant MOVIE (PI: Marcella Marconi) and the PRIN MUR 2022 project (code 2022ARWP9C) ``Early Formation and Evolution of Bulge and HalO (EFEBHO)", PI: Marcella Marconi, funded by the European Union -- Next Generation EU. VBT acknowledges financial support from the Indian Space Research Organisation (ISRO), Department of Space, Government of India, through the ISRO Research Project (Proposal No. RAC-S/GU/2024/4/22). VBT also acknowledges partial support from the IUCAA Associateship Programme.

\section*{Data Availability}
To ensure full reproducibility of our stellar evolution calculations, we have made the complete set of \texttt{MESA} inlist files used in this work publicly available. These inlists specify all relevant physical assumptions and numerical controls, including the adopted initial masses, metallicity, mixing-length parameters, atmospheric boundary conditions, mesh refinement settings, and nuclear reaction network options.

The repository contains the master inlist as well as the auxiliary inlists required to reproduce the full grid of stellar models discussed in this paper. All models were computed using the publicly released version of \texttt{MESA} described in \citet{paxton2019}. No additional, unpublished modifications to the source code were applied. The inlist files are available at \href{https://github.com/mrinmaymedhi/mesa-hd20794\string#GitHub}{GitHub}.

The data used in the manuscript can be obtained upon reasonable request from the corresponding author.




\bibliographystyle{mnras}
\bibliography{references} 



\appendix
\section{Complete Nuclear Network Output}
This appendix provides supplementary material related to the extended nuclear reaction network used in the \texttt{MESA} simulations of HD~20794. The full network includes isotopes up to chlorine and tracks their temporal evolution self-consistently throughout the stellar lifetime. The additional tables and figures demonstrate that, aside from hydrogen and helium burning, no significant nuclear processing occurs for elements beyond the CNO group in HD~20794. In particular, the abundances of phosphorus and chlorine remain constant in both the core and envelope, consistent with the absence of activated proton-capture channels under the physical conditions encountered in this low-mass, metal-poor star.

\begin{table}
\centering
\caption{Complete surface abundance output from the extended nuclear network for the best-fit HD~20794 model (age $\approx 9$~Gyr). This table includes short-lived radioactive isotopes and species with negligible surface abundances, listed for completeness only.}
\label{tab:abundances_all_hd20794}
\setlength{\tabcolsep}{1pt}
\begin{tabular}{clccccc}
\hline
Sl. No. & Isotope & Mass Fraction & $\epsilon_{\mathrm{model}}$
& $\epsilon_{\mathrm{lit}}$ & $\Delta(\epsilon)$ & Reference \\
\hline
1  & $^{1}$H   & $7.425\times10^{-1}$ & 12.00 & 11.96 & +0.04 & -- \\
2  & $^{2}$H   & $1.520\times10^{-41}$ & -28.99 & -- & -- & -- \\
3  & $^{3}$He  & $2.650\times10^{-5}$  & 7.07 & -- & -- & -- \\
4  & $^{4}$He  & $2.550\times10^{-1}$  & 10.92 & 11.50 & -0.58 & -- \\
5  & $^{7}$Li  & $2.800\times10^{-9}$  & 2.73 & 3.65 & -0.92 & \cite{cretignier2023} \\
6  & $^{7}$Be  & $1.000\times10^{-95}$ & $< -80$ & -- & -- & -- \\
7  & $^{8}$B   & $1.010\times10^{-99}$ & $< -80$ & -- & -- & -- \\
8  & $^{12}$C  & $7.662\times10^{-4}$  & 7.93 & 8.43 & -0.50 & \cite{mortier2013} \\
9  & $^{13}$C  & $9.300\times10^{-6}$  & 5.98 & -- & -- & \cite{mortier2013} \\
10 & $^{13}$N  & $1.010\times10^{-99}$ & $< -80$ & -- & -- & -- \\
11 & $^{14}$N  & $2.264\times10^{-4}$  & 7.34 & 8.05 & -0.71 & -- \\
12 & $^{15}$N  & $8.920\times10^{-7}$  & 4.90 & -- & -- & -- \\
13 & $^{14}$O  & $1.010\times10^{-99}$ & $< -80$ & -- & -- & -- \\
14 & $^{15}$O  & $1.010\times10^{-99}$ & $< -80$ & -- & -- & -- \\
15 & $^{16}$O  & $2.105\times10^{-3}$  & 8.25 & 8.77 & -0.52 & \cite{mortier2013} \\
16 & $^{17}$O  & $8.340\times10^{-7}$  & 4.82 & -- & -- & -- \\
17 & $^{18}$O  & $4.750\times10^{-6}$  & 5.55 & -- & -- & -- \\
18 & $^{17}$F  & $1.010\times10^{-99}$ & $< -80$ & -- & -- & -- \\
19 & $^{18}$F  & $1.010\times10^{-99}$ & $< -80$ & -- & -- & -- \\
20 & $^{19}$F  & $1.120\times10^{-7}$  & 3.90 & -- & -- & -- \\
21 & $^{18}$Ne & $1.010\times10^{-99}$ & $< -80$ & -- & -- & -- \\
22 & $^{19}$Ne & $1.010\times10^{-99}$ & $< -80$ & -- & -- & -- \\
23 & $^{20}$Ne & $4.340\times10^{-4}$  & 7.46 & 8.05 & -0.59 & \cite{asplund2021} \\
24 & $^{22}$Ne & $3.510\times10^{-5}$  & 6.33 & -- & -- & -- \\
25 & $^{22}$Mg & $1.010\times10^{-99}$ & $< -80$ & -- & -- & -- \\
26 & $^{24}$Mg & $1.780\times10^{-4}$  & 7.00 & 7.54 & -0.54 & \cite{mortier2013} \\
27 & $^{31}$P  & $2.190\times10^{-6}$  & 4.98 & 5.50 & -0.52 & \cite{maas2017} \\
28 & $^{32}$S  & $9.890\times10^{-5}$  & 6.62 & 7.10 & -0.48 & \cite{reddy2003} \\
29 & $^{33}$Cl & $1.010\times10^{-99}$ & $< -80$ & -- & -- & -- \\
30 & $^{34}$Cl & $1.010\times10^{-99}$ & $< -80$ & -- & -- & -- \\
31 & $^{35}$Cl & $1.300\times10^{-6}$  & 4.70 & 5.38 & -0.68 & -- \\
32 & $^{34}$Ar & $1.010\times10^{-99}$ & $< -80$ & -- & -- & -- \\
33 & $^{35}$Ar & $6.369\times10^{-4}$  & 7.39 & -- & -- & -- \\
\hline
\end{tabular}
\end{table}

\section{Helium Abundance Sensitivity}
\label{sec:helium}
To assess the impact of helium enrichment on the inferred stellar parameters, we computed evolutionary sequences spanning helium mass fractions corresponding to a standard Galactic enrichment relation \citep{Peimbert1974},
\begin{equation}
 Y = Y_p + \left(\frac{\Delta Y}{\Delta Z}\right) Z ,   
\end{equation}
with a primordial helium abundance, $Y_p$ = 0.245 \citep{Cyburt2016}, and helium-to-metal enrichment ratio, $\Delta Y/\Delta Z = 1.5-2.5$ \citep{Balser2006}, at fixed metallicity $Z = 0.005$. This range yields helium abundances $Y = 0.2525-0.2575$. Each sequence was evolved to an age of 10 Gyr, appropriate for HD~20794, and the resulting global stellar parameters were compared. 

\begin{table}
\centering
\caption{Helium sensitivity test at 9 Gyr for a $0.8\,M_\odot$ model with $Z=0.005$ and $\alpha_{\rm MLT}=1.4$.}
\label{tab:helium_sensitivity}
\begin{tabular}{lcccccc}
\hline
$Y$ & $X$ & $T_{\rm eff}$ (K) & $L/L_\odot$ & $R/R_\odot$ & $\log g$ \\
\hline
0.2525 & 0.7425 & 5479.9 & 0.6942 & 0.9244 & 4.409 \\
0.2550 & 0.7400 & 5495.2 & 0.7071 & 0.9277 & 4.406 \\
0.2575 & 0.7375 & 5512.3 & 0.7245 & 0.9333 & 4.401 \\
\hline
\end{tabular}
\end{table}

Across the explored helium range, the maximum variation at 9 Gyr is 
$\Delta T_{\rm eff} \lesssim 33$~K,
$\Delta L/L_\odot \approx 1.9\times10^{-2}$~dex,
$\Delta \log g \approx 8\times10^{-3}$~dex, and
$\Delta R \approx 8.9\times10^{-3}\,R_\odot$. These shifts are comparable to or smaller than the observational uncertainties and do not alter the qualitative behaviour of the models or the principal conclusions of this work. In particular, the global structure and surface abundance evolution remain unaffected at a level relevant for the present analysis.

\section{Overshooting Sensitivity Test}
\label{sec:overshoot}
To assess the impact of convective overshooting on the derived stellar parameters, we performed additional evolutionary calculations including mild envelope overshooting using the exponential prescription implemented in \texttt{MESA}. For a $0.8\,M_\odot$ main-sequence star such as HD~20794, the core remains radiative throughout its evolution, and therefore convective core overshooting is not physically relevant \citep{Kippenhahn2012,Salaris2005}. 

\begin{table}
\centering
\caption{Overshooting sensitivity test at 9 Gyr for a $0.8\,M_\odot$ model with $Z=0.005$ and $\alpha_{\rm MLT}=1.4$.}
\label{tab:overshoot_sensitivity}
\begin{tabular}{lcccc}
\hline
Model & $T_{\rm eff}$ (K) & $L/L_\odot$ & $R/R_\odot$ & $\log g$ \\
\hline
No overshoot        & 5474.14 & 0.6853 & 0.9204 & 4.413 \\
$f_{ov} = 0.01$          & 5460.67 & 0.6741 & 0.9173 & 4.416 \\
$f_{ov} = 0.02$          & 5460.67 & 0.6741 & 0.9173 & 4.416 \\
\hline
\end{tabular}
\end{table}

We instead tested envelope overshooting at the base of the convective envelope using values $f_{\rm ov} = 0.01$ and $f_{ov} = 0.02$ \citep{paxton2013}. Each model was evolved to an age of 10 Gyr and compared with the baseline (no-overshoot) model. Relative to the baseline model, the maximum variations at 9 Gyr are 
$\Delta T_{\rm eff} \approx 13$~K,
$\Delta L/L_\odot \approx 7\times10^{-3}$~dex,
$\Delta \log g \approx 3\times10^{-3}$~dex, and
$\Delta R \approx 3.1\times10^{-3}\,R_\odot$. These differences are substantially smaller than the observational uncertainties and do not modify the inferred evolutionary state or surface abundance conclusions. As expected for a low-mass star with a radiative core, envelope overshooting has a negligible impact on the global stellar structure.

\bsp	
\label{lastpage}
\end{document}